\documentstyle[epsfig]{mn}

\newif\ifAMStwofonts

\ifoldfss
  \ifCUPmtlplainloaded \else
    \NewTextAlphabet{textbfit} {cmbxti10} {}
    \NewTextAlphabet{textbfss} {cmssbx10} {}
    \NewMathAlphabet{mathbfit} {cmbxti10} {} 
    \NewMathAlphabet{mathbfss} {cmssbx10} {} 
  \fi
  \ifAMStwofonts
    \ifCUPmtlplainloaded \else
      \NewSymbolFont{upmath} {eurm10}
      \NewSymbolFont{AMSa} {msam10}
      \NewMathSymbol{\upi}     {0}{upmath}{19}
      \NewMathSymbol{\umu}     {0}{upmath}{16}
      \NewMathSymbol{\upartial}{0}{upmath}{40}
      \NewMathSymbol{\leqslant}{3}{AMSa}{36}
      \NewMathSymbol{\geqslant}{3}{AMSa}{3E}

       \let\le=\leqslant
       \let\ge=\geqslant
    \fi
  \fi
\fi 

\ifnfssone
  \newmathalphabet{\mathit}
  \addtoversion{normal}{\mathit}{cmr}{m}{it}
  \addtoversion{bold}{\mathit}{cmr}{bx}{it}
  \newmathalphabet{\mathbfit} 
  \addtoversion{normal}{\mathbfit}{cmr}{bx}{it}
  \addtoversion{bold}{\mathbfit}{cmr}{bx}{it}
  \newmathalphabet{\mathbfss} 
  \addtoversion{normal}{\mathbfss}{cmss}{bx}{n}
  \addtoversion{bold}{\mathbfss}{cmss}{bx}{n}
  \ifAMStwofonts
    \ifCUPmtlplainloaded \else
      \UseAMStwoboldmath
      \makeatletter
      \new@mathgroup\upmath@group
      \define@mathgroup\mv@normal\upmath@group{eur}{m}{n}
      \define@mathgroup\mv@bold\upmath@group{eur}{b}{n}
      \edef\UPM{\hexnumber\upmath@group}
      \new@mathgroup\amsa@group
      \define@mathgroup\mv@normal\amsa@group{msa}{m}{n}
      \define@mathgroup\mv@bold\amsa@group{msa}{m}{n}
      \edef\AMSa{\hexnumber\amsa@group}
      \makeatother
      \mathchardef\upi="0\UPM19
      \mathchardef\umu="0\UPM16
      \mathchardef\upartial="0\UPM40
      \mathchardef\leqslant="3\AMSa36
      \mathchardef\geqslant="3\AMSa3E

       \let\le=\leqslant
       \let\ge=\geqslant
    \fi
  \fi
\fi 

\ifnfsstwo
  \DeclareMathAlphabet{\mathbfit}{OT1}{cmr}{bx}{it}
  \SetMathAlphabet\mathbfit{bold}{OT1}{cmr}{bx}{it}
  \DeclareMathAlphabet{\mathbfss}{OT1}{cmss}{bx}{n}
  \SetMathAlphabet\mathbfss{bold}{OT1}{cmss}{bx}{n}
  \ifAMStwofonts
    \ifCUPmtlplainloaded \else
      \DeclareSymbolFont{UPM}{U}{eur}{m}{n}
      \SetSymbolFont{UPM}{bold}{U}{eur}{b}{n}
      \DeclareSymbolFont{AMSa}{U}{msa}{m}{n}
      \DeclareMathSymbol{\upi}{0}{UPM}{"19}
      \DeclareMathSymbol{\umu}{0}{UPM}{"16}
      \DeclareMathSymbol{\upartial}{0}{UPM}{"40}
      \DeclareMathSymbol{\leqslant}{3}{AMSa}{"36}
      \DeclareMathSymbol{\geqslant}{3}{AMSa}{"3E}

       \let\le=\leqslant
       \let\ge=\geqslant
    \fi
  \fi
\fi 

\ifCUPmtlplainloaded \else
  \ifAMStwofonts \else 
    \def\upi{\pi}
    \def\umu{\mu}
    \def\upartial{\partial}
  \fi
\fi

\def\gsim{\lower.73ex\hbox{$\sim$}\llap{\raise.4ex\hbox{$>$}}$\,$}
\def\lsim{\lower.73ex\hbox{$\sim$}\llap{\raise.4ex\hbox{$<$}}$\,$}
\def\mpc{$\,h^{-1}\,$Mpc}

\def\%{~per~cent}

\title{The Local Hole in the Galaxy Distribution: Evidence from 2MASS}
\author[W.J. Frith et al.]
{W.J.~Frith, G.S.~Busswell, R.~Fong, N.~Metcalfe \& T.~Shanks \\
Department of Physics, University of Durham, Science Laboratories, South Road, Durham, DH1 3LE, United Kingdom\\}

\pagerange{\pageref{firstpage}--\pageref{lastpage}}
\pubyear{2003}

\begin{document}

\maketitle

\label{firstpage}

\begin{abstract}
Using the bright galaxy counts from the 2 Micron All Sky Survey (2MASS) second incremental
release, two techniques for probing large-scale structure at
distances of $\sim~150$\mpc\ are investigated. First, we study
the counts from two sets of six $5^\circ\times\sim~80^\circ$ strips
in the two galactic caps. In the six southern
strips a deficit of $\sim30$\% was found relative to a predicted
homogeneous distribution at $K_{s}\sim11$. These strips were also in good agreement
with a model incorporating  the radial density function of the
southern 2dF Galaxy Redshift Survey (2dFGRS), which shows a
deep underdensity between $\sim90$ and 180\mpc. Together with a similar
underdensity found in the Las Campanas Redshift Survey, these
results indicate a very large `local hole' in the Southern Galactic Cap (SGC) to
\gsim150\mpc\ with a linear size across the sky of $\sim200$\mpc\, 
but with a significantly lower mean underdensity of
$\sim$30\% than that suggested by the bright APM SGC counts.
The counts in the northern set of strips are low overall but indicate a more
varied pattern. When all the available 2MASS
data with $|b|>30^\circ$ were aggregated, they
indicated underdensities of $\sim18$\% and $\sim30$\% at $K_{s}\sim11$
for the northern and southern areas respectively. Our second method
uses the ratio of the counts with $11.38<K_{s}<12.38$ to
$12.88<K_{s}<13.38$ binned in 25 deg$^2$ areas; the counts from these areas 
provide a smoothed map over the sky of the slope in the counts
between $K_{s}$=11.38 and~13.38. Visually, the resulting map shows
the expected complex form of the cosmic web and picks out known rich
clusters, demonstrating the usefulness of this `slope statistic' as
a probe of large-scale structure at distances of $\lsim150$\mpc. Most
interestingly, the map also shows large regions, $\sim100^\circ$
across, of steep counts in both hemispheres. Thus, the present 2MASS
data suggest the presence of a potentially huge contiguous void
stretching from south to north. Not only would this delineate
further the limits for the Cosmological Principle but it would also
show the possible presence of significant power on scales of
\gsim300\mpc\ in the galaxy power spectrum.

\end{abstract}


\section{Introduction}
Over the last twenty years, our understanding of the galaxy
distribution in the local universe has advanced dramatically. The
observational success since the 1920s of Friedmann universes
reflects a fundamental aspect of the Universe, its isotropy and
homogeneity, expressed theoretically by the Cosmological Principle.
Indeed, it is in the context of Friedmann universes that recent
observations of the cosmic microwave background, SNIa and galaxy
redshift surveys provide evidence for a non-zero cosmological
constant. On the other hand, the observations of large-scale
structure in the galaxy distribution shows that on even quite large
scales the Universe is highly inhomogeneous, with the galaxy
distribution tracing out a cosmic web of filamentary and wall-like
structures.

Thus, the question arises as to the scales on which the Universe can
be viewed as isotropic and homogeneous, i.e., the scales at which
the Cosmological Principle can be said to hold. 
Recent work on the local distribution of galaxies indicates
the existence of `great walls' of $\sim100$\mpc\ in size
and $\sim30$--50\mpc\ voids (e.g. Bellanger \& De Lapparent, 1995).
These observations would seem to support the possibility that the
steeper than expected APM $B$-band galaxy counts over 4300 deg$^2$
of the Southern Galactic Cap (SGC) might indeed reflect a very large local hole in
the Universe (Shanks 1990). But for our present understanding of the
cosmic web, as gathered from dynamical N-body simulations, it would
be unexpectedly large and deep (Baugh et al., 2002). Such a local hole would clearly
place stringent constraints on any model of large-scale structure.
However, using CCD photometry Metcalfe, Fong \&~Shanks (1995) found
a small residual scale error in the APM zero-points for $B$\gsim17.
The corrected APM counts fainter than $B\sim17.5$ were now in
good agreement with the expected counts from normal homogeneous
models; a deficit of $\sim50$\% at $B\sim15.5$ still remained along with
the possibility of a very large deep hole in the SGC.

The presence of an underdensity in the SGC is now being corroborated
by preliminary results from deep galaxy redshift
surveys, probing the galaxy distribution to
$z\sim0.2$. The number-redshift histograms, $n(z)$, of the Southern
surveys display astonishing structures, with large regions of
underdensity at $z\sim0.03$--0.06 bounded by
sharp peaks of overdensity (Zucca et al., 1997; Shectman et al.
1996; Ratcliffe et al., 1996; Norberg et al., 2002).

Interestingly, the 2 Micron All Sky Survey (2MASS; Jarrett et al., 2000) is in
the process of completing an infrared survey of the whole sky. Even
with the preliminary data released by 2MASS it is possible to test
the observational results indicating a large hole. In particular,
compared with the photgraphic APM survey, a survey based on digital
photometry has fewer concerns over errors in the calibration and
field-to-field variations in the zero-point and limiting magnitude.
However, a complete interpretation of the 2MASS infrared counts is
at present hindered by uncertainties in the surveyed area arising
from the incompleteness of the survey and an intricate mask not yet
publicly available.
 
With the goal of determining the approximate variations in the
galaxy distribution over the galactic caps, and the scale of local
large-scale inhomogeneities, we have thus compared the $K_{s}$-band
2MASS counts from several declination strips over the galactic caps
with models incorporating the radial density functions of the
2dF Galaxy Redshift Survey (2dFGRS; Colless et al., 2001), in
their northern and southern areas. To see if the features seen in these
strips persisted over larger solid angles we then examined the combined
available counts for $|b|\ge$30$^\circ$. For a more
detailed view of these inhomogeneities, we used variations in the
slope of the 2MASS counts between $K_{s}=11.38$ and 13.38 as a probe
of inhomogeneities on scales of $\sim5^\circ$. In Section~2 we
present general details of the data samples used. In Section~3 we first
examine the optical magnitude counts from the APM survey and 2dFGRS, and then
carry out the aforementioned analyses of the 2MASS data. The results are summarised
and discussed in Section~4.

\section{The Data}
\subsection{The APM Survey}

\begin{figure}
\begin{center}
\centerline{\epsfxsize = 3in
\epsfbox{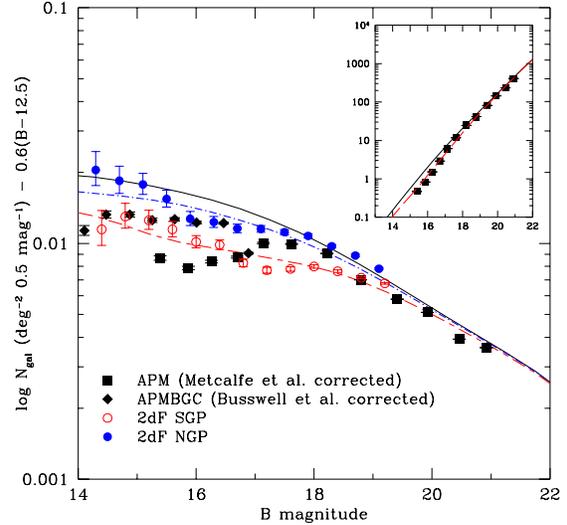}}
\caption{Magnitude counts from the Metcalfe et al. corrected APM survey (Maddox et al., 1990a,b),   
the Busswell et al. corrected APMBGC and 2dF parent catalogue counts for
the northern and southern fields (Norberg - priv. comm.);
the errorbars indicated are Poissonian. The main
diagram indicates the counts with a subtracted Euclidean slope; the inset
shows the Metcalfe et al. corrected APM counts on an ordinary logarithmic
scale for reference. The continuous line represents
the expected homogeneous trend, while the dashed and dot-dashed lines show the
models derived from the 2dF radial density profiles for $B\le17.5$ in the southern and
northern fields respectively as calculated from Figs. \ref{fig:z2dfs} and
\ref{fig:z2dfn}.}
\label{fig:apm}
\end{center}
\end{figure}

\begin{figure}
\begin{center}
\centerline{\epsfxsize = 3in
\epsfbox{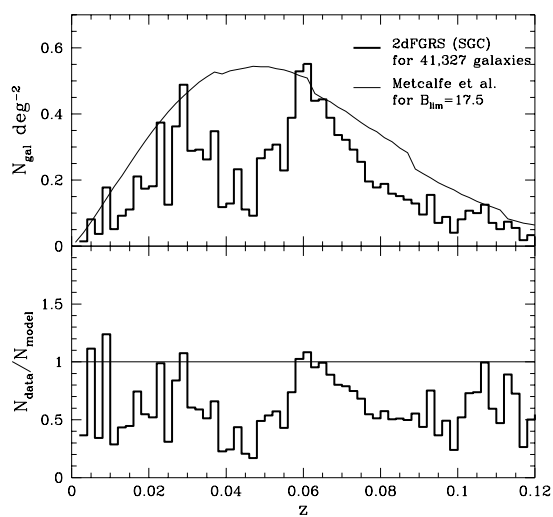}}
\caption{The galaxy redshift distribution for 41,327 galaxies from the 2dFGRS
SGC declination strip. The top figure shows the number of galaxies per redshift bin
per square degree; the data is normalised such that the ratio of the
number of galaxies predicted by the model to the observed number matches the
corresponding ratio in the 2dF SGP counts shown in fig.~\ref{fig:apm}.
The continuous line indicates the expected trend
calculated from the PLE model detailed in section 3.1 for $B_{lim}=17.5$, using a 
blanket colour correction of $B-b_{j}$=0.2 to transfer the 2dF parent catalogue to the $B$-band.
The bottom figure shows the radial density profile for this data.} 
\label{fig:z2dfs}
\end{center}
\end{figure}

\begin{figure}
\begin{center}
\centerline{\epsfxsize = 3in
\epsfbox{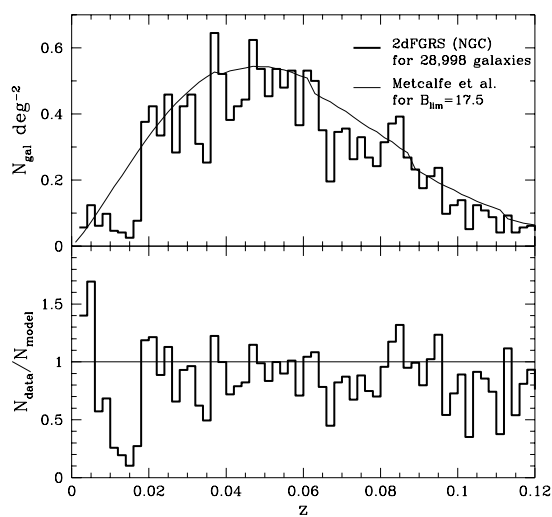}}
\caption{The galaxy redshift distribution for 28,998 galaxies in the
2dFGRS NGC equatorial field. Again the top figure shows the number of galaxies per redshift bin
per square degree; the data has been normalised in the same way as the southern distribution.
The continuous line indicates the expected trend for $B_{lim}$=17.5, again using the blanket $B-b_{j}$=0.2
colour correction but with an additional zero-point correction of 0.1 mag. found by Busswell et al. (2002) for
the northern field with respect to their CTIO northern photometry; the residual $B-b_{j}$ correction to the 2dF
parent catalogue is 0.1 mag. The bottom figure shows the radial density profile for this data.}
\label{fig:z2dfn}
\end{center}
\end{figure}

The APM survey is the largest complete galaxy survey to date,
covering a contiguous $\sim$4300~deg$^2$ area in the SGC, with 20
million images brighter than $b_{j}=22$. It is based on the APM photometry of
185 photographic Schmidt plates in the region $\alpha\sim21^h$ to
$5^h$ and $\delta\sim-70^\circ$ to $-20^\circ$ (Maddox et al.,~1990a,b). As 
mentioned in section~1, the Metcalfe et al. (1995) corrections to the APM photometry 
are used here with a mean colour correction of $B-b_{j}=0.2$. 

We have also used the publicly available APM Bright Galaxy Catalogue (APMBGC; Loveday et al., 1996),
which reaches a magnitude limit of $b_{j}=16.44$. Using very accurate photometric
data taken from the CTIO over the approximate current 2dFGRS declination strips,
a zero-point correction of 0.313 magnitudes, found from over 700 matched galaxies, 
has been applied to these counts (Busswell et al., 2002).  

\subsection{The 2dF Galaxy Redshift Survey}

The 2dFGRS is selected in the photographic $b_{j}$ band using the 
APM survey and subsequent alterations and extensions to it (Norberg et al., 
2002) for two declination strips in the northern and southern galactic caps, as
well as 99 randomly selected $2^\circ$ fields scattered over the southern APM field. 
The data used in this paper is taken from the 100k data release detailed in 
Colless et al. (2001), which has made spectroscopic and photometric data publicly available 
for $\sim90,000$ galaxies in the SGC and NGC, 
bounded approximately by $\alpha\sim22^h$ to
$3^h30^m$ and $\delta\sim-32^\circ$ to $-26^\circ$ in the southern declination strip and 
$\alpha\sim10^h20^m$ to $13^h40^m$ and $\delta\sim-6^\circ$ to $0^\circ$ in the NGC. 
Despite the fact that the original 2dFGRS galaxy sample was selected with an extinction-corrected 
limit of $b_{j}=19.45$, subsequent linearity and zero-point recalibrations of the 2dF parent catalogue, 
as well as revised extinction estimates over the 2dF fields, result in median limiting magnitudes of 
$b_{j}=19.40$ over the SGC field, and $b_{j}=19.35$ in the NGC field. The 2dF galaxy data used in this paper has 
been selected with a quality flag of $Q\ge3$ corresponding to a redshift 
reliability of $>$90\% for 92.7\% of the total sampled galaxies (Colless et al. 2001).  

\subsection{The 2MASS Extended Source Catalogue}

The 2 Micron All Sky Survey (2MASS) is a digital survey of the whole sky
in the $K_s$, $J$ and $H$-bands. Through the 2MASS second incremental release
the photometry for $\sim6\times10^5$ galaxies is publicly available
with a complete detection rate to $K_s=13.5$ (Jarrett et al., 2000). 

The $K_{s}$-band magnitudes used in this paper have been determined
using the method of Cole et al. (2001). In a comparison with the
deeper $K$ photometry of Loveday (2000), Cole et al. found that the
most accurate $K_s$ Kron magnitudes are obtained from
the 2MASS $J$-band Kron magnitudes, colour corrected using the $J-K$
default aperture colours, with a zero-point offset of 0.061~mag. An
additional correction is also needed to convert these Kron
magnitudes to total magnitudes; a blanket 0.12~mag correction was
applied to account for this and the zero-point offset. A full
account of these corrections can be found in Cole et al.

\begin{figure*}
\begin{center}
\centerline{\epsfxsize = 5.5in
\epsfbox{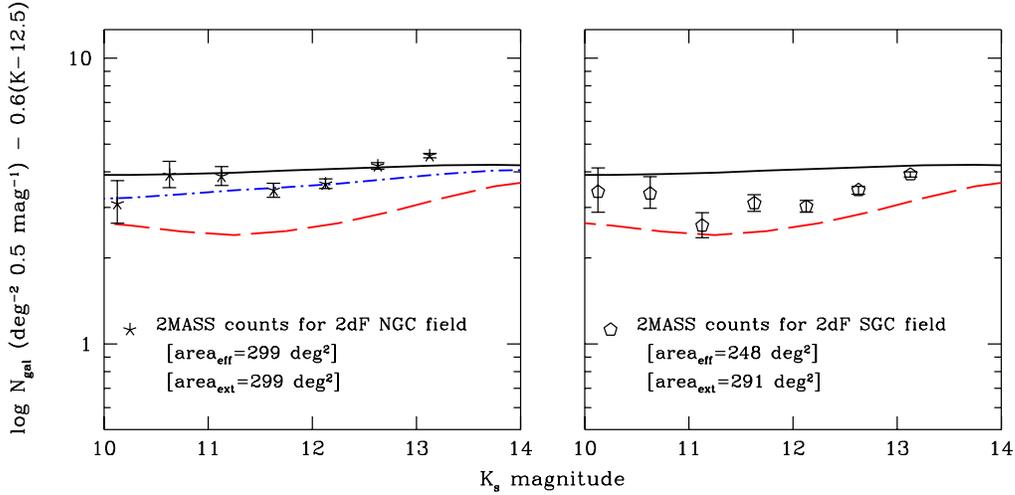}}
\caption{Galaxy magnitude counts extracted from 2MASS for the
approximate current 2dF declination fields. The appropriate $K$-band models are indicated,
with the homogeneous counts indicated by the solid line, the 2dF SGC
model with the dashed line and the 2dF NGC model by a dot-dashed line as in fig.~\ref{fig:apm}.
The effective and extracted areas are also shown, corresponding to the solid angle of the selected field and
the calculated solid angle accounting for gaps in the data.}
\label{fig:m2df}
\end{center}
\end{figure*}

\section{Bright Galaxy Counts}

\subsection{Optical galaxy counts}

The originally published APM galaxy counts for the SGC show a
steeper than expected slope brightward of $B\sim19$ (Maddox et
al., 1990). Subsequent studies by Metcalfe et al. (1995) have
revealed a scale error in the original APM photometry for $17<B<19$,
yielding counts faintward of $B=18$ which are more consistent with a homogeneous 
model for the universe. However, the galaxy counts are still very low brightward
of $B\sim18$ (with a galaxy count of $\sim50$\% with respect to the homogeneous prediction at $B\sim15.5$~mag.), and
the photometry here is uncertain. However new photometry checks of the APMBGC (Loveday et al., 1996)
with very accurate CTIO photometry over the approximate 2dF declination fields (Busswell et al., 2002)
indicate a 0.313 mag. zero-point correction to the bright photometry
yielding APM counts for $B<17$ which are higher
but still significantly underdense (a galaxy count of $\sim$75\% with respect to the homogeneous
prediction at $B\sim15.6$~mag.). 
Fig.~\ref{fig:apm} shows the APM SGC counts 
with the Metcalfe et al. corrections, and the Busswell et al. corrected APMBGC counts using a colour correction
to the B-band of $B-b_{j}=0.2$. Similar comparisons of the Busswell et al. CTIO photometry with the 2dF parent catalogue
indicate a mean 0.1 magnitude zero-point error over the northern field such that the
2dF magnitudes are too faint, and a negligible
off-set in the south; the corrected 2dF parent catalogue counts (Norberg - priv. comm.)
are also shown using the same $B-b_{j}$ colour correction. The models shown in Fig.~\ref{fig:apm} 
are based on the Pure Luminosity Evolution (PLE) model of Metcalfe et al. (2001) incorporating a 
$K+E$-correction detailed in Bruzual \& Charlot (1993). 

Recent redshift surveys are beginning to elucidate the cause of
the steeper than expected slope of bright galaxy counts.
Fig.~\ref{fig:z2dfs} and Fig. \ref{fig:z2dfn} show the 2dFGRS redshift distributions in the 
upper panels and the radial density profiles below 
for the 100k data release (Colless et al., 2001) for the southern 
and northern declination fields respectively. We have imposed a limiting magnitude of $B_{lim}$=17.5
in order to more accurately match the 2MASS limiting K-band magnitude and therefore probe a 
similar redshift range. 
The 2dF data are compared to an expected homogeneous distribution using the Metcalfe et 
al. (2001) PLE prediction. In order to avoid uncertainties in the effective area and redshift completeness
the data has been normalised such that the ratio of predicted to observed galaxies is equal
to the ratio of the corresponding cumulative 2dF counts to the cumulative number of galaxies 
predicted by the homogeneous number count model for $B\le17.5$. 
Most striking in the southern distribution is the 
visible large-scale structure, characterised by deficiencies  
at 0.03\lsim$z$\lsim0.06 ($\sim$90 to 180\mpc) and
$z\lsim0.06$z$\lsim0.1$ ($\sim$180 to 300\mpc). The NGC field
also indicates a deficiency below z$\sim0.02$, beyond which the distribution
is approximately homogeneous.

We first wish to check if the observed 2dF galaxy distributions in
the SGC and NGC might explain the low southern galaxy-magnitude
counts, and the apparent asymmetry between the north and south. Incorporating the radial 
density profiles from the 2dF areas (the ratio of the observed distribution to
the expected homogeneous counts from Figs.~\ref{fig:z2dfs} and \ref{fig:z2dfn}) into the homogeneous model as a 
coefficient of the luminosity function parameter $\phi^*$, yields an
expected trend for the corresponding galaxy-magnitude counts. For z$>$0.12 a constant,
homogeneous $\phi^*$ is used). 
The dot-dashed and dashed lines in Fig.~\ref{fig:apm} show the predictions
of these variable $\phi^*$ models for the northern and southern 2dF counts
respectively. The observed 2dF counts and the corresponding
models show excellent agreement. 

The low trends of the 2dF SGC counts and the APM 
B$>$17 and APMBGC B$<$17 counts imply significant structure
especially since the APM
southern field covers $\sim4300$~deg$^2$. The presence of significant
large-scale structure is also indicated in the Southern $n(z)$ of
the Las Campanas Redshift Survey (Shectman et al. 1996), which
samples an area of $7.5^\circ\times80^\circ$ in the SGC 
$\sim15^\circ$ southward of the 2dF area. The 2dF NGC variable $\phi^*$ model and 2dF NGC counts are low 
but indicate a much smaller deficiency in the counts. This can 
be observed in the 2dF NGC $n(z)$ (Fig.~\ref{fig:z2dfn}) which is closer
to the homogeneous prediction. 

This apparent asymmetry in the SGC and NGC 
$n(z)$ and the scale and distance of
the SGC inhomogeneities are intriguing, and can be further
investigated using the presently released 2MASS data. 

\subsection{2MASS galaxy counts}

\begin{figure*}
\begin{center}
\centerline{\epsfxsize = 5.5in
\epsfbox{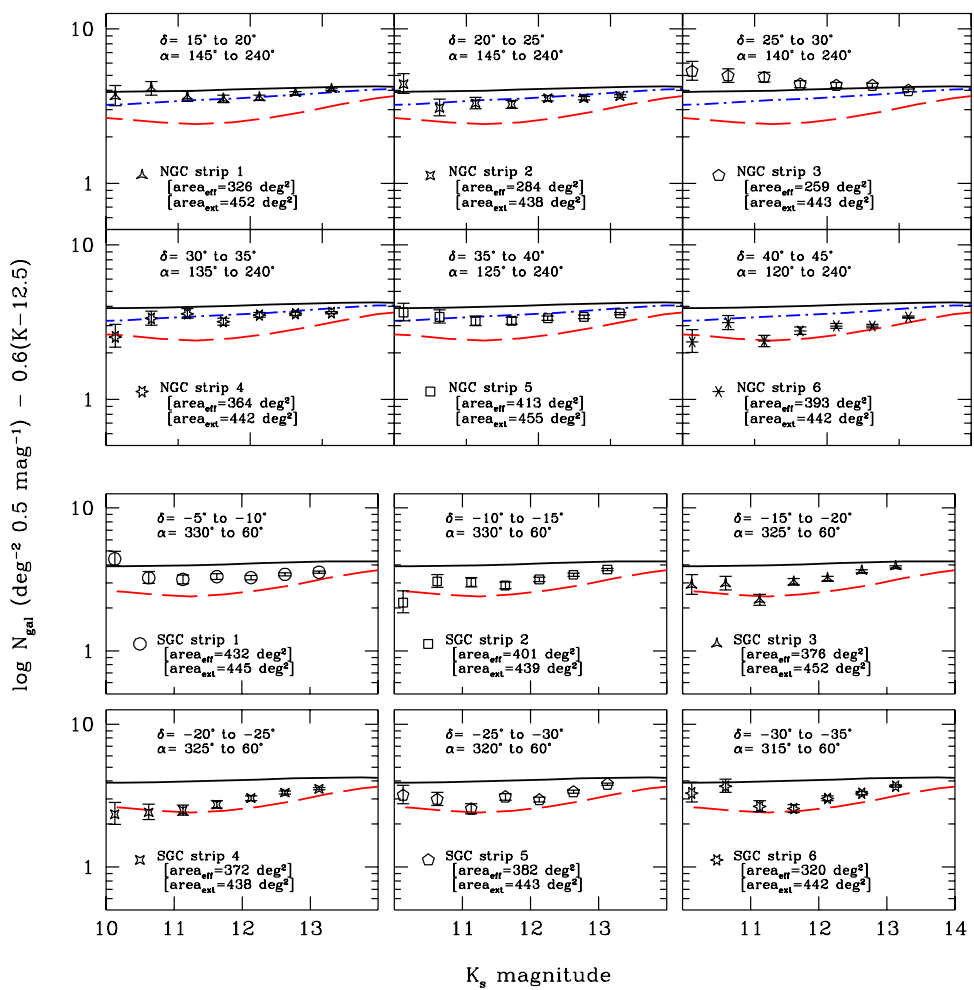}}
\caption{Magnitude counts extracted from 2MASS in declination strips 
each 5 degrees wide from the NGC and SGC, shown in the top and bottom panels
respectively. Again the counts are shown with the Euclidean slope
subtracted and with Poisson errors. The homogeneous PLE model is
indicated by the continuous line, the model derived from the
southern 2dF radial density profile for $B\le17.5$ is shown in all strips by a
dashed line, and the northern model is indicated in the northern
strips by a dot-dashed line. The effective and extracted areas are also
indicated for each strip.} 
\label{fig:strip}
\end{center}  
\end{figure*}

\begin{figure*}
\begin{center}
\centerline{\epsfxsize = 5.5in
\epsfbox{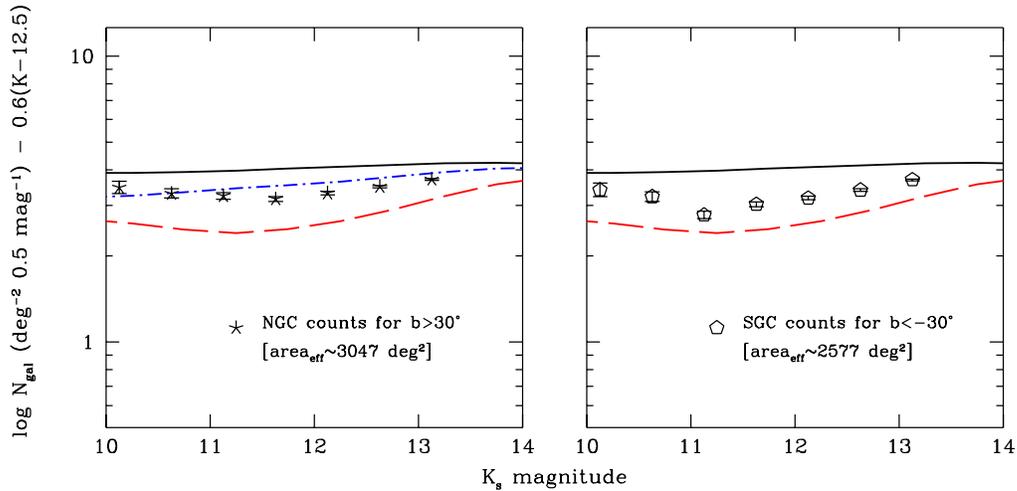}}
\caption{Galaxy magnitude counts extracted from 2MASS for the northern ($b\ge30^\circ$)
and southern ($b\le-30^\circ$) galactic caps. The counts here are normalised to the mean number
seen in the corresponding set of strips at $K_{s}=13.13$. The appropriate models are
indicated as in fig. ~\ref{fig:m2df}. The effective areas indicated here correspond to the
effective solid angle inferred by the normalisation.}
\label{fig:caps}
\end{center}
\end{figure*}

\begin{figure*}
\begin{center}
\centerline{\epsfxsize = 6.5in
\epsfbox{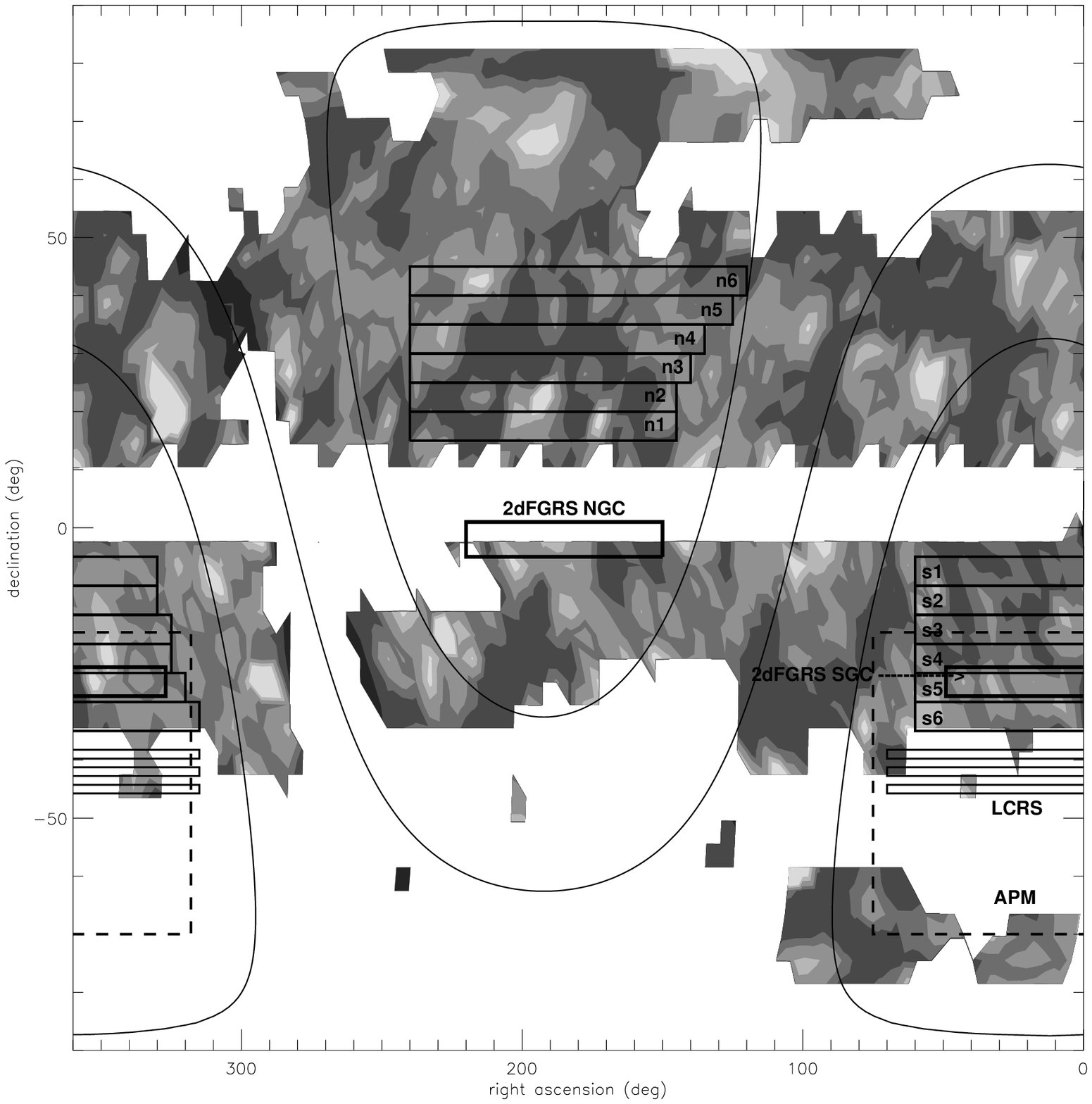}}
\caption{The ratio of 2MASS galaxies in the ranges 11.38$<K_{s}<$12.38
and 12.88$<K_{s}<$13.38 taken for 25 deg$^2$ areas as a function of
position on the sky. The least shaded regions indicate shallow counts
corresponding to a galaxy distribution of $<50$\% with respect to the homogenous
prediction, with progressively darker areas indicating
density levels of $<60$\%, $<80$\%, $<100$\%, $<200$\% and $>200$\%; these contour
levels are used to emphasise the underdense regions of interest. Also shown are the
2MASS strip and galaxy survey fields of interest.
Galactic longitude lines are shown for $b=-30^\circ$, $0^\circ$ and $30^\circ$.
For reference, the Coma galaxy cluster
is at $\alpha=195^\circ$, $\delta=+28^\circ$ and the Shapley cluster is
at $\alpha=202^\circ$, $\delta=-27^\circ$.}
\label{fig:map}
\end{center}
\end{figure*}

The asymmetry observed in the SGC and NGC optical magnitude counts and 2dFGRS n(z) can be further
investigated using the 2MASS second incremental release. Firstly, we wish to examine the trend of 
the K-band counts for the 2dF fields to determine whether the observed deficiencies and north-south
asymmetry is present in the K-band, and also to verify the usefulness
of the variable $\phi^*$ models described in Section 3.1. 
Fig.~\ref{fig:m2df} shows the 2MASS $K_{s}$ counts extracted from the 
approximate 2dF 100k data release fields for which 2MASS data is available. Also shown are the homogeneous 
prediction and the variable $\phi^*$ models for the 2dFGRS SGC and NGC
fields using the radial density function for $B\le17.5$ (approximately matching the
limiting $K_{s}$ magnitude of 2MASS and therefore more likely to produce a 
better description of the counts). 

Unfortunately, the 2MASS mask for the second incremental release  
is fairly complex and the effective areas had to be determined
approximately. This was carried out by searching for gaps in the 2MASS
astrometry of over 0.5 degrees on the right ascension axis in 0.5 degree
declination strips; the maximum right ascension range of any missing
area was then determined. The declination range of unobserved areas
was similarly inferred yielding an upper limit on the area of the
missing region. A visual inspection of the distribution of the 2MASS
galaxies on the sky reveals that this estimate of the effective area
is likely to be reasonably good for fairly complete strips.

The resulting $K_{s}$ counts are in excellent agreement with
the $B$-band 2dF counts; comparing the counts at approximately equivalent
magnitudes, the northern counts indicate deficiencies
of 0.0\% (at $B=15.1$) and 2.7\% (at $K_{s}=11.13$), and the southern counts
deficiencies of 28\% (at $B=15.2$) and 32\% (at $K_{s}=11.13$). The $K_{s}$ counts
for the 2dF fields are also in approximate agreement with the predictions
of the corresponding variable $\phi^*$ models.

Together with the structure seen in the 2dF $n(z)$, these 2MASS
$K_{s}$ counts show that such bright counts in areas as small as
$\sim300$~deg$^2$ can be used as a diagnostic probe of large
underdensities at distances of $\sim150$\mpc\ since the slope of the counts between
$K_{s}\sim$11 and $K_{s}\sim$13 is dependent on the redshift distribution at these 
distances; small fluctuations in the galaxy
distribution at greater distances have little effect on the number counts
at these magnitudes (Frith 2001). 

Using this comparison between bright counts and the 2dF variable $\phi^*$ models,
it is now possible to investigate the angular extent of the
local hole in the SGC using 2MASS. To this end, the counts were determined for
six 5 degree declination strips in the SGC (Fig.~\ref{fig:strip}), each with approximately
the same solid angle and right ascension range as the current 2dF
fields (see Fig.~\ref{fig:m2df}). In order to determine the robustness of the north--south asymmetry 
observed in the 2dF survey and the corresponding $B$ and 
$K_{s}$ band counts, 2MASS counts were also
taken for another six similar strips in the NGC. The number and positions of
these strips were restricted firstly by the constraints of the mask
of the available 2MASS data, and secondly by regions of significant 
extinction; declinations with $|b|\ge 30^\circ$ were selected. 
 
The southern strip counts in Fig.~\ref{fig:strip} are in approximate agreement with the 2dF SGC variable 
$\phi^*$ model with a uniform steep slope in the counts over the entire
declination range, similar to the 2MASS counts for the 2dF SGC field (Fig.~\ref{fig:m2df}),
indicating that the `local hole' 
extends as far north as $\delta = 0^\circ$ in the SGP, and as
far south as can be sampled with the present 2MASS catalogue.
Considering that the LCRS $n(z)$ (Shectman et al. 1996) indicates a similar
underdensity at $\sim150$\mpc, $\sim15^\circ$ south
of the 2dF SGP field, it would seem to
support the presence of a `local hole' in the SGC of at least $\sim80^\circ\times40^\circ$ in
extent ($\sim65$\% of the APM SGC area). 

In the northern strips, the counts in Fig.~\ref{fig:strip} are more varied with approximate
agreement with the 2dF NGC varaiable $\phi^*$ model and the form of the 2MASS counts
for the 2dF NGC field in all but two strips.
The shallow counts in the third strip ($25^\circ < \delta
< 30^\circ$) are probably due to the presence of the Coma cluster and
its associated supercluster. 

In the strip counts of both hemispheres,
similar mean deficiencies for the southern and northern aggregated counts
at $K_{s}=13.13$ of $\sim12$\% are found corresponding to a mean depth 
of $\ge150\,h^{-1}\,$Mpc, possibly indicating that the inhomogeneities 
in the local galaxy distribution exist on larger scales. 

It is worth noting that the slopes of the counts
in the northern strips in Fig.~\ref{fig:strip} are fairly close to Euclidean, and that a renormalisation
of the model or a 10\% error in the effective areas would compensate for the 
apparent deficiency. However, these compensation effects are likely to be fairly small since,
firstly, any change in the model normalisation would affect the excellent agreement the model has with 
fainter K-band data (e.g. Gardner et al., 1996; Huang et al., 1996; Glazebrook et al., 1995),
and secondly, since any significant recalculation of the effective areas would upset the excellent
agreement of the 2dF optical magnitude counts with the 2MASS counts for the 2dF fields (Fig.~\ref{fig:m2df}) 
which are normalised using the same method as the two sets of six 2MASS strips in Fig.~\ref{fig:strip}.   

\subsection{$|b|\ge30^\circ$ 2MASS galaxy counts}

To investigate the local galaxy distribution further, we have extended the
area studied from the two sets of six strips to an area over the
galactic caps to $|b|\ge30^\circ$ (see Fig.~\ref{fig:caps}). Moving to 
these larger areas, we see that the hole in the SGC persists with a 
deficiency of $\sim30$\% with respect to the homogeneous model at $K_{s}=11.13$,
similar to those seen in the SGC strips. In the NGC, the approximate homogeneous trend observed
in the B$\le$17.5 2dF NGP n(z) is replaced by counts which are underdense by $\sim18$\% at $K_{s}=11.13$, 
indicating the possible presence of a large void in the NGC as well as the SGC. It is interesting to 
note also that the counts in both hemispheres display comparable deficiencies of
$\sim$20\% at $K_{s}$=11.63. Clearly as the northern area gets bigger, the picture gets more complicated
with areas of underdensity also appearing as in the SGC.

Due to the complexity of the mask for the
2MASS second incremental release, we have found it difficult to estimate the actual
solid angle surveyed and have simply normalized the counts in Fig.~\ref{fig:caps} to the
mean number density observed in the corresponding set of strips in Fig. ~\ref{fig:strip} at the 
$K_{s}=13.13$ half magnitude bin; this is likely to be a reasonable procedure since the 
implied solid angle of the 2MASS mask over the galactic caps is only 
a factor of $\sim2$ larger than the combined effective area of the strips. 

Thus, the north--south asymmetry observed in the 2dFGRS n(z) and optical magnitude counts
and the 2MASS counts selected for 2dF declination strips 
appears to be in disagreement with the 2MASS counts over the
galactic caps. A possible explanation comes from a more detailed analysis over 
larger areas provided in the next section. 

\subsection{Mapping the Local Galaxy Distribution}

Finally, we wish to investigate the form and extent of the local hole at a
higher resolution and over a larger area than the $|b|\ge30^\circ$ areas
while avoiding calculations of the effective area. It is clear from
the 2MASS counts and the variable $\phi^*$ models that an essential signature for a local
hole at $\sim150$\mpc\ distance is the steep slope between
$K_{s}\sim11$ and $\sim13$. We have therefore examined the 2MASS galaxy
counts over the entire available area
in this magnitude range binned in 25 deg$^2$ areas with a constant 
declination range of $5^\circ$. With
the distances being explored at these magnitude limits, this amounts
to a smoothing on a linear scale of $\sim10$\mpc. For each area we
then calculated the ratio of the number of galaxies in the whole
magnitude bin $11.38<K_{s}<12.38$, to that in the half-magnitude bin
$12.88<K_{s}<13.38$, a procedure in which the two numbers are similar
and the error on the ratio reasonable. Finally, we divided this ratio by
the predicted ratio of the homogeneous model. This `slope statistic'
is then a measure of how steep or shallow the counts are relative to
the homogenous slope. Equivalently, if we assume the counts in
25 deg$^2$ areas are approximately homogeneous by $K_{s}=13.13$, the
ratio is then just the observed count in the $K_{s}\sim11.88$ half magnitude bin as a fraction of
that predicted by the homogeneous model; it then indicates the
extent to which the area is underdense or overdense at distances of
$\lsim150$\mpc. Clearly, for this latter interpretation of the `slope
statistic' there is some error, as the smoothing area is an order of
magnitude smaller than the declination strips of the previous
section and so a uniform number density in the $K_{s}=13.13$ bin, as is observed
in the strips, cannot be assumed. However, it may at least prove useful as a qualitative
probe of local density inhomogeneities. Fig.~\ref{fig:map} is a
contour map of this statistic.

A particular advantage of this technique is that it is independent
of an accurate determination of the effective area, while allowing
us to probe the galaxy distribution over large contiguous areas of sky
in the volume below $\sim150$\mpc.
However, galaxy magnitudes at low galactic latitudes of
$|b|<30^\circ$ begin to be effected by significant extinction even in the $K$-band,
and the mapped statistic in these regions is unreliable for the
purposes here. 

The use of the map in Fig.~\ref{fig:map} as an indicator of
inhomogeneities in the distribution of galaxies on the sky is
supported by, first, its ability to pick out known clusters 
at distances of $\lsim150$\mpc; the Coma cluster and its
associated supercluster can clearly be seen; the Shapley cluster
also appears as a high density area. Secondly, we also see generally good agreement with
the results in the previous section for the declination strips. This
is typified by the more patchy form of the galaxy distribution in
the NGC than in the SGC, where overdense regions exist around the pole
but accidentally cancel with surrounding underdensities to produce low counts in four of the
NGC strips. 

From Fig.~\ref{fig:map} it can also be observed that the very large `local hole' that we have
been exploring in the SGC is a coherent aggregation of smaller
holes. In particular, we see that the 2dF SGC area lies in a general
region of such holes or voids, with a remarkably large area of right  
ascension between $280^\circ$ and $50^\circ$ and declination between
$-2^\circ$ and $-35^\circ$.

Fig.~\ref{fig:map} also supports the relative
similarity of the NGC and SGC cap counts for $|b|>30^\circ$ with
large regions of shallow counts in the NGC countered only by the area
of steeper counts at the NGP. It is also now apparent why the counts in the 
northern strip 3 have a much flatter slope than the southern strips
and the northern and southern cap counts, since there appears to be a particularly 
dense region in the galaxy distribution over the NGP, surrounded by a region of much 
shallower counts, further supporting the possibility of a large void in the NGC.

Fig.~\ref{fig:map} also indicates that the galaxy distribution is
highly varied as might be expected with the cosmic web smoothed on
scales of $\sim10$\mpc. In particular, individual deep holes,
underdense by as much as $\gsim50$\%, with sizes of
$\sim10$--$20^\circ$ can also be observed. Considering
the depth of the survey this corresponds to a linear scale of
$\sim20$--60\mpc, similar to the scale of voids found using large
redshift surveys (e.g. Bellanger \& De Lapparent, 1995).

\section{Summary and Conclusions}

In this paper, we first reviewed the evidence from APM and 2dFGRS
optical number-magnitude counts that the B$<$18 mag. counts
particularly in the SGC are steeper than expected on the basis of
simple homogeneous models. One possible interpretation is that this is
caused by a large local hole in the distribution of galaxies. However,
other possible interpretations are that the effect is caused by
evolution or poor photographic photometry.  New data, principally the
2dFGRS $n(z)$ and the 2MASS second incremental release, have therefore
been used to discriminate between these various possibilities. With
the 2dFGRS, any local hole can be sought in the redshift distribution
directly and the 2MASS $K_{s}$ counts can be used as a check of the steepness of
local galaxy counts in a band where a linear detector has been used and
the effects of evolution are likely to be less than in the $B$-band.

The 2MASS data is not in its final form and there are problems in
particular with an uncertain mask in various areas of the
survey. We therefore first made a detailed analysis of the 2MASS data
in the NGC and SGC 2dFGRS fields each of $\sim$300 deg$^2$ area to
check for the hole in the south and investigate the previous apparent
north--south asymmetry in the counts.

Then we extended this approach to a wider area of the NGC and SGC,
first using strip counts in $\sim$1800deg$^2$ areas near each of the
galactic poles in each hemisphere, then overall NGC
and SGC counts based on all available 2MASS areas with
$|b|\ge$30$^\circ$. Finally, using a 2MASS count `slope statistic', we
made a higher resolution ($\sim$10\mpc) map to investigate the
detailed form of the large-scale structure to an average depth of
$\sim$150\mpc, again over the full available 2MASS area.

The results are summarised as follows:-

\begin{itemize}

\item{The 2MASS $K_{s}$ number count in the 2dFGRS SGC field is
significantly steeper than in the 2dFGRS NGC field, consistent with
previous results in the $B$-band.}

\item{Large holes are seen in the galaxy distribution in the 2dFGRS SGC $B\le17.5$ 
$n(z)$ out to z$<$0.1 which are not seen in the NGC 2dFGRS $B\le17.5$ $n(z)$.}

\item{Variable $\phi^*$ number count models which track the 2dFGRS 
$n(z)$'s give a good fit to the $K_{s}$-band galaxy 
counts in both the NGC and SGC 2dFGRS areas, suggesting that the
steep SGC counts are due to the holes in the 2dFGRS SGC $n(z)$.}

\item{The six 2MASS SGC strip counts over an $\sim$2000 deg$^2$ area 
consistently show the same underdensity as seen in the 2dFGRS SGC
field. Counts over a similar area in the NGC show a more varied
behaviour with some strips being higher than even the homogeneous model 
and others starting to show the same level of underdensity as seen in the
SGP. }

\item{2MASS $K_{s}$ counts over the whole SGC and NGC areas with
$|b|\ge$30$^\circ$ indicate an underdensity in the SGC which is only slightly
less than in the 2dFGRS SGC field and an increased underdensity in
the NGC area.}

\item{Examining the 2MASS counts at a higher resolution
with the `slope statistic' reveals a more patchy form
to the galaxy distribution in the NGC than in the SGC. In particular, 
overdensities associated with the Coma and Shapley clusters can clearly be
detected. These areas are surrounded by regions
of underdensity resulting in the reasonably homogeneous
counts seen in the 2dFGRS NGC field and the six NGC strips. The SGC shows
more evidence of contiguous underdensity again consistent with the steep
$K_{s}$ counts seen in the 2dFGRS SGC field and the six SGC strip counts.}

\item{Using the evidence from 2MASS using the `slope statistic' and
the six SGC strips, and also the underdensity seen in the 
LCRS (Shectman et al. 1996), we conclude
that an underdensity of $\sim$30\% in the SGC extends over at least
280$^\circ<\alpha<~$50$^\circ$ and -45$^{\circ}<\delta<0^{\circ}$ to
$\gsim150$\mpc. This corresponds to a linear size across the sky of
$\sim$200\mpc.}

\item{The `slope statistic' map also supports the form of the $b>30^\circ$
counts, indicating large regions of underdensity in the NGC away from
the NGP and moderated only by the effect of the Coma cluster at
the pole, indicating the possible presence of a huge, connected local
underdensity present in both galactic caps of scale \gsim300\mpc.}

\end{itemize}

\section*{Acknowledgments}
This publication makes use of data products from the Two Micron
All Sky Survey, which is a joint project of the University of 
Massachusetts and the Infrared Processing and Analysis Centre/California
Institute of Technology, funded by the Aeronautics and Space Administration
and the National Science Foundation. We thank 
Tom Jarrett for his helpful information 
on the 2MASS mask. We would also like to thank Carlton
Baugh and Shaun Cole for useful discussions, and Richard Bower
and Phil Outram for technical help.

\label{lastpage}

\end{document}